\documentclass[10pt,twocolumn]{article} %% two column, final layout
\usepackage{ol2}
\usepackage[draft]{hyperref}
\usepackage{amsmath}

\begin{document}

\twocolumn[ %% activate for two-column option

\title{A new type of light with enhanced optical chirality}

\author{Carmelo Rosales-Guzm\'an $^{1}$, Karen Volke-Sepulveda ${^2}$ and Juan P. Torres $^{1,3}$}

\address{$^1$ ICFO-Institut de Ciencies Fotoniques, 08860 Castelldefels, Barcelona, Spain \\
$^2$ Instituto de Fisica, UNAM, Apdo. Postal 20-364, 01000 Mexico D.F., Mexico  \\
$^3$ Dep. of Signal Theory and Communications, Univ. Politecnica
de Catalunya, 08034, Barcelona, Spain}

\email{juanp.torres@icfo.es}

\begin{abstract}
Recently, Tang and Cohen (Science \textbf{332}, 333, 2011) have
demonstrated a scheme to enhance the chiral response of molecules,
which relies on the use of circularly polarized light in a
standing wave configuration. Here we show a new type of light
which possesses orbital angular momentum and enhanced chiral
response. Unexpectedly, in the locations where the beams show enhanced
optical chirality, only the longitudinal components of the
electric and magnetic fields survive, showing a new way to yield
optical chirality different from the usual.
\end{abstract}


] %% activate for two-column option

\noindent

Chirality plays a crucial role in life, since most of the
important molecular building blocks of life, i.e. aminoacids and
sugars, come in left- or right-handed varieties. One of the most
striking features of life is why most of these molecules present
a specific chirality at all, since many chemical processes
performed in the lab to obtain these substances give no preference
for any specific form of chirality \cite{anderson1972}.

An object is chiral if it cannot be superimposed with its own
mirror image. A pair of such chiral objects or systems are called
enantiomorphs, or enantiomers for the special case of molecules.
Enantiomers are identical in most regards, it is only in their
interaction with other chiral objects that they become
distinguishable \cite{barron2004}. Circularly polarized light
(CPL) is an example of chiral object, and there is a myriad of
optical phenomena, usually referred to as {\em optical activity},
whose origin can be reduced to the different response of molecules
to right- and left-circularly polarized light.

One example of these effects is magnetic optical rotation, a
rotation of the plane of polarization of a linearly polarized beam
when it propagates in a chiral medium, which can be described in
terms of different refractive indices for the two types of CPL.
Another example is circular dichroism (CD), the different
absorption rate of chiral molecules under the presence of left-
and right-circularly polarized light, which translates in the
transformation of a linearly polarized beam into an elliptically
polarized one. Both phenomena coexist in frequency regions which
present absorption.

The different rate of absorption of a chiral medium when
illuminated by the two forms of CPL is generally small, which can
make its detection rather demanding in some cases. Up to recently,
it was thought that this response depends only on intrinsic
properties of the chiral medium. But recently, a pseudoscalar
quantity termed {\em optical chirality} ($C$) was introduced by
Yiqiao and Cohen \cite{tang2010} to quantify the amount of
chirality present in an arbitrarily-shaped optical field.
Crucially, the inspection of this quantity shows that it should be
possible to generate superchiral fields \cite{hendry2010,tang2011}
to produce an enhancement of the amount of circular dichroism
detected. This opens a whole new scenario for the detection of
optical chirality, where now the shape of the optical field plays
a crucial role for enhancing the detection of chirality.

Light with orbital angular momentum (OAM) is also a chiral object.
Light beams with OAM show an azimuthal phase dependence in the
transverse plane of the form $\sim \exp \left( i m
\varphi\right)$, where the index $m$, which can take any integer
value, determines the OAM of the beam and $\varphi$ is the
azimuthal angle in cylindrical coordinates. In general, the spin
(polarization) and orbital contributions to the total angular momentum cannot be considered
separately \cite{jackson1}. However, in the paraxial regime,
both contributions can be manipulated independently
\cite{allen-book}.

In spite of being chiral objects, all experimental efforts aimed
at detecting a chiral response of molecules making use of optical
beams with OAM have apparently failed
\cite{araoka2005,loffler2011}. Some theoretical investigations of
the interaction of beams with OAM and molecular systems have
yielded seemingly contradictory results. Within the electronic
dipole approximation for diatomic molecules, and in the paraxial
approximation for light beams, it was argued \cite{babiker2002}
that the internal ``electronic-type" motion does not participate in
any OAM change, while later on, the inclusion of electronic,
rotational, vibrational and center-of-mass motion variables,
seemed to demonstrate \cite{alexandrescu2005} that the OAM can
couple to the rotational and electronic motion. Again under the
paraxial approximation, it was established that OAM cannot be
engaged with the chirality of a molecular system
\cite{andrews2004}.

Here we will show that certain types of optical beams endowed with
OAM can indeed present an enhanced local chiral response, even
larger than the response that would be obtained with the usual
circularly-polarized light, so in principle, they might be used to
detect an enhanced circular dichroism effect. For this purpose, we
make use of two basic ingredients. First, we consider a form of
light-matter interaction \cite{barron2004,craig1984} that couples
the electric and magnetic fields of the optical beam to the
electric $\mathbf{p}$ and magnetic $\mathbf{m}$ dipole moments of
a chiral molecule, so that
\begin{eqnarray}
& & \mathbf{p}=\mu_E \mathbf{E}+i G \mathbf{B} \nonumber \\
& & \mathbf{m}=\mu_B \mathbf{B}-i G \mathbf{E},
\label{constitutive_relations}
\end{eqnarray}
where $\mu_E$, $\mu_B$ and $G$  are the electric, magnetic and
electric-magnetic dipole polarizabilities, respectively.  Even
though higher-order multipoles can also contribute for light beams
with general spatial shapes \cite{yang2011}, we assume here that
these contributions are sufficiently small so they can be safely
neglected. Secondly, we consider Bessel light beams endowed with
OAM \cite{karen2002}. Bessel beams are exact solutions of Maxwell
equation's, and as we will see below, this departure from the
usual case of the paraxial regime allows to unveil some important
features not easily shown in the paraxial framework.

The time-averaged optical chirality of a CW beam of the form
${\cal E}(\mathbf{r},t)=1/2\, \mathbf{E} ({\bf r}) \exp(i k_z
z-i\omega t) + h.c.$,  where $\omega$ is the angular frequency and
$k_z$ is the longitudinal component of the wave vector, writes
\cite{tang2010}
\begin{equation}
C=\frac{\omega \epsilon_0}{2} \Im \left[\mathbf{E} ({\bf
r})\cdot\mathbf{B}^{*}({\bf r}) \right]. \label{chirality}
\end{equation}
$\Im$ stands for the imaginary part, $\epsilon_0$ is the
permittivity of vacuum, $\mathbf{E}$ and $\mathbf{B} $ are the
electric and magnetic fields, respectively, and ${\bf r}$ is the
position. For circularly polarized light with a polarization
vector of the form $\hat{x}+\sigma i\hat{y}$ ($\sigma=\pm 1$), the
optical chirality is $C_{CPL}=2\sigma k U_e$, where
$U_e=\epsilon_0 |\mathbf{E}|^2/4$ is the local average energy
density of the field and $k$ is the wavenumber of the light beam.

In particular, we consider a Bessel beam which propagates along
the $z$ direction of the form \cite{karen2002}

\begin{eqnarray}
& & \mathbf{E}(\mathbf{r}) = E_0 \left\{ \left( \alpha
\hat{\mathbf x}+\beta \hat{\mathbf y} \right)
J_m(k_t\rho)\exp\left( im \varphi
\right)  \right. \nonumber \\
& &  + i \frac{k_t}{2k_z} \left[ \left(\alpha+i\beta \right)
J_{m-1}(k_t\rho)\exp\left[ i(m-1)\varphi \right] \right. \nonumber \\
& & \left. \left. -\left(\alpha-i\beta
\right)J_{m+1}(k_t\rho)\exp\left[ i(m+1)\varphi \right] \right]
\right\}\hat{\mathbf z}, \label{electric}
\end{eqnarray}
where $J_m$ is the $m$th-order Bessel function, $\alpha$ and
$\beta$ are complex constants indicating the polarization state,
($\rho,\varphi$) are the radial and azimuthal variables in
cylindrical coordinates, $m$ is the winding number related to the OAM of the Bessel beams and $k_t$ is the
transverse component of the wave vector satisfying $k=\sqrt{k_t^2+k_z^2}$.

The magnetic field can be computed from Eq. (\ref{electric})
through Maxwell's equations. Here we are interested in the field
at the center of the beam ($\rho=0$) for the cases $m=\pm 1$. Even
though these are vortex beams possessing OAM, the total electric
and magnetic fields at the center do not vanish; only the
transverse components vanish, but the longitudinal components of
the fields survive. We consider the coherent superposition, with
complex weights $A$ and $B$, of two OAM optical beams with indices
$m=+1$ and $m=-1$, and equal linear polarizations given by $\cos
\phi \hat{\mathbf{x}}+\sin \phi \hat{\mathbf{y}}$, i.e.,
$\alpha=\cos \phi$ and $\beta=\sin \phi$ in Eq. (\ref{electric}).

The electric and magnetic fields at the center write
\begin{eqnarray}
& & \mathbf{E}(0) = i\frac{E_0 k_t}{2k_z} \left[A\exp \left( i\phi
\right)-B \exp \left(- i\phi
\right) \right] \hat{\mathbf{z}}\nonumber \\
& &  \mathbf{B}(0) = -\frac{E_0k_t}{2\omega} \left[A\exp \left(
i\phi \right)+B \exp \left(- i\phi \right) \right]
\hat{\mathbf{z}} \label{fields_at_center}
\end{eqnarray}
Inserting the expressions of $\mathbf{E}(0)$ and $\mathbf{B}(0)$
into Eq. (\ref{chirality}), the energy density $U_e$ and the
optical chirality $C$ turns out to be
\begin{eqnarray}
& & U_e=\frac{\epsilon_0  k_t^2|E_0|^2 }{k_z^2} \left[
|A|^2+|B|^2-2|A||B| \cos \left( 2 \phi-\xi\right) \right]
\nonumber \\
& & C=-\frac{\epsilon_0 k_t^2 |E_0|^2}{k_z} \left( |A|^2-|B|^2
\right) \label{chiral}
\end{eqnarray}
where $\xi=\arg(B/A)$ is the phase difference between the complex
weights $A$ and $B$. Again, the optical chirality does not
generally vanish at the center of the beam ($\rho=0$), what is no
longer true for all other cases with $m \ne \pm 1$.

The structure of the electric and magnetic fields which bear
optical chirality is radically different from the usual form of
circularly-polarized light. The fields at the center of the beam
contain a single component of the field (the longitudinal
component along the direction of propagation $\hat z$), while in
the case of circularly-polarized light, there are two orthogonal
components, $\hat x$ and $\hat y$, perpendicular to the direction
of propagation of the beam. For instance, for $\phi=0$, and $A$
and $B$ real numbers, Eq. (\ref{fields_at_center}) shows that
there is a $\pi/2$ phase difference between the electric and
magnetic fields, which is responsible for the non-zero value of
the chirality. This $\pi/2$ phase difference is also typical of
circular polarized light.

One of the manifestations of the presence of optical chirality is
the detection of circular dichroism, which is usually quantified
by means of the dissymmetry factor,
$g=2(A^{+}-A^{-})/(A^{+}+A^{-})$, where $A^{\pm}$ is the
absorption rate of chiral molecules when illuminated with chiral
fields of opposite chirality. For the type of interaction
considered in Eq. (\ref{constitutive_relations}), the relative
dissymmetry factor writes \cite{tang2010} $g/g_{CPL}=C/(2k U_e)$,
where $g_{CPL}$ is the dissymmetry that would be measured with a
circular polarized beam.

\begin{figure}[!t]
\centering 
\includegraphics[height=7cm]{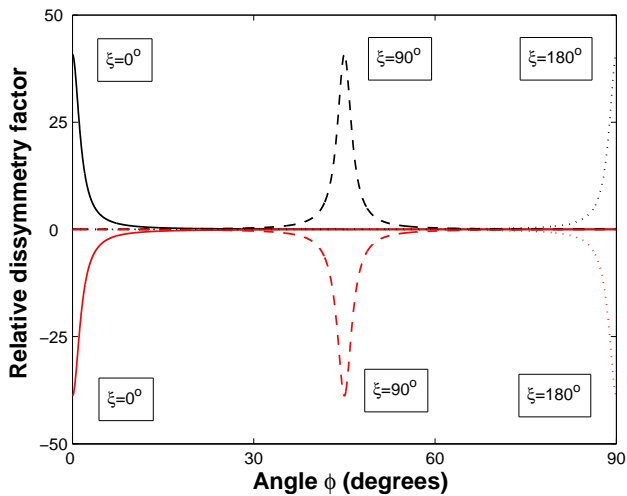}
\caption{(Color online) Relative dissymmetry factor $g/g_{CPL}$ as a function of
the polarization angle $\phi$ for two values of the ratio
$r=|B|/|A|$: $r=0.95$ (red lines) and $r=1.05$ (black lines), and
three values of the angle $\xi$: $0^o$ (solid), $90^o$ (dashed)
and $180^o$ (dotted). In all cases $k_t/k=0.1$.} \label{figure1}
\end{figure}

For a molecule located at the center of the Bessel beam, one
easily obtains
\begin{equation}
\frac{g}{g_{CPL}}=-\frac{k_z}{k}
\frac{|A|^2-|B|^2}{|A|^2+|B|^2-2|A||B|\cos \left(2\phi+\xi
\right)}, \label{dissymmetry}
\end{equation}
Inspection of Eq. (\ref{dissymmetry}) shows that when $|A|=0$ or
$|B|=0$, the dissymmetry factor for paraxial beams ($k_z \sim k$)
is nearly that of circularly polarized light, i.e. $|g/g_{CPL}|
\sim 1$. And moreover, by choosing appropriate values of $A$ and
$B$, so that the energy density at the center of the beam is close
to zero, one can obtain a large value of the dissymmetry factor,
i.e. $|g/g_{CPL}| \gg 1$, the key towards observing an enhanced
chiral response \cite{tang2010}.

Fig. \ref{figure1} shows the relative dissymmetry factor as a
function of the polarization angle ($\phi=0$ corresponds to
polarization along $\hat{x}$, while $\phi=90^o$ corresponds to
polarization along $\hat{y}$), for some selected values of the
angle $\xi$ and the ratio $r=|B|/|A|$. For $r=0.95$, one obtains
$g/g_{CPL}<0$, while for $r=1.05$, one has $g/g_{CPL}>0$. Eq.
(\ref{dissymmetry}) shows that an enhanced chiral response
requires $|A| \sim |B|$, and $\phi=\xi/2$, so that the relative
dissymmetry factor reach the maximum value of $g/g_{CPL}=(k_z/k)
(|A|+|B|)/(|B|-|A|)$. The case $|A|=|B|$ would generate a null of
the total electric field at the center.

Two important conclusions can be drawn from Eq.
(\ref{dissymmetry}). Firstly, we can detect, in principle, the
circular dichroism induced by a chiral molecule located at the
center of a Bessel beam with winding numbers $m=\pm 1$. This
is somehow unexpected, since the optical field at the center of
the beam contains a single component of the electric and magnetic
fields, and the transverse fields vanish. Second, the circular
dichroism can even be largely enhanced when compared with the case
of circularly-polarized light, similarly to the effects observed
in \cite{tang2011} with counter-propagating circularly polarized
beams. Furthermore, we want to stress that in contrast with the
latter case, the superposition proposed here involves two
co-propagating fields, whose centers coincide along the
propagation axis, avoiding the experimental problem of locating
the sample at one node of an standing wave \cite{tang2011}.

The circular dichroism considered here could be experimentally
observed by using as probe a single molecule with a fixed
absorption dipole moment parallel to the beam axis
\cite{novotny2001} or a chiral solid microsphere\cite{Cipparrone2011}, located at the center of the beam. The
fluorescence of the single molecule can probe the local field
intensity before and after the interaction of the light beam with
the chiral medium. The probe particle can be trapped at the center of
the Bessel beam by means of an auxiliary gaussian-like beam ({\em optical tweezer}) or by direct optical trapping with the Bessel beam itself.

In conclusion, we have unveiled a new type of light-matter
interaction to obtain an enhanced chiral response. It makes use of
light beams with OAM, and surprisingly, the electric and magnetic
fields do not present the usual form corresponding to circularly
polarized light. Notice that in the interaction of a molecule with
the electric and magnetic fields, only the local fields at the
position of the molecule are of interest. The approach taken here
is thus in contrast to the one consider in
\cite{bliokh2011,andrews2012}, where the total optical chirality
of the beam is considered. In that case, the optical chirality,
when integrated over the whole beam, yields the global unbalance
of the two spin ($\sigma= \pm 1$) angular momentum components.

This work was supported by projects FIS2010-14831 and PHORBITECH
(FET-Open 255914), and by Fundacio Privada Cellex Barcelona. K.
V-S acknowledges support from CONACYT Mexico (grant 132527). We
thank A. E. Cohen for useful and illuminating conversations about
the general issue of light-matter interactions in chiral media.

\end{document}